\documentclass[12pt,preprint,floatfix,aps]{revtex4}
\usepackage{graphicx,float}

\begin{document}
\begin{center}
{\Large \bf 
Anomalous Diffusion in View of Einstein's
1905 Theory of Brownian Motion}

\vspace{0.5cm}

{\large 
Sumiyoshi Abe$^{1,*}$ and Stefan Thurner$^{2,**}$}

{\small 
  $^{1}${\it Institute of Physics, University of Tsukuba, Ibaraki 305-8571, Japan} \\
  $^{2}${\it Complex Systems Research Group, HNO, Medizinische Universit\"at Wien,} \\
{\it W\"ahringer G\"urtel 18-20, Vienna A-1090, Austria}  \\}  
\vspace{0.8cm}
\end{center}

\begin{center}
{\small \bf Abstract}
\end{center}

\noindent
{\small Einstein's theory of Brownian motion is revisited in order to formulate generalized kinetic theory 
of anomalous diffusion. It is shown that if the assumptions of analyticity and the existence of the second
moment of the displacement distribution are relaxed, the fractional derivative naturally appears in the
diffusion equation. This is the first demonstration of the physical origin of the fractional derivative, in marked contrast to the usual phenomenological introduction of it.
Furthermore, Einstein's approach is generalized to nonlinear kinetic theory to derive the porous-medium-type 
equation by the appropriate use of the escort distribution.	
							
\vspace{1cm}
 }

\vfill
\noindent
{\footnotesize
$^*$ suabe@sf6.so-net.ne.jp\\
 $^{**}$ thurner@univie.ac.at
}  

\newpage

   In his landmark 1905 paper \cite{einstein05}, Einstein has presented a new derivation of the diffusion equation. In this derivation, there appear two kinds of distributions. One is the probability, $f(x, t)dx$, of finding a Brownian particle in an interval $[x,x+dx]$ at time $t$, and the other is the probability density, $\phi(\Delta)$ for displacement, $\Delta$, of the particle within a single discrete time step, $\tau$. The basic evolution equation is the following:
 \begin{equation}
 f(x,t+\tau) dx = dx \int_{-\infty}^{\infty} d\Delta f(x+\Delta,t) \phi(\Delta) \quad ,
 \end{equation}  
where $\phi(\Delta)$ satisfies the condition, $\phi(\Delta)=\phi(-\Delta)$. It is clear that this equation is consistent with the normalization conditions on $f(x, t)$ and $\phi(\Delta)$. Assuming analyticity of $f(x, t)$ in terms of both $x$ and $t$ and existence of the second moment of $\phi(\Delta)$, Einstein derived the diffusion equation, $\partial f / \partial t= D \partial^2 f / \partial x^2$, where the diffusion constant is given by
 \begin{equation}
 D=\frac{1}{\tau} \int_{-\infty}^{\infty} d\Delta \frac{\Delta^2}{2} \phi(\Delta) \quad .
 \end{equation}  
As well known, the mean displacement accordingly grows in time as
 \begin{equation}
 \sqrt{\bar{x^2}} = \sqrt{2Dt} \quad ,
 \end{equation}  
where the over-bar denotes the expectation value with respect to $f(x, t)$. It is important to note that Eq.~(1) is different from today's familiar method of Green's function, 
$f(x,t+\tau)=\int_{-\infty}^{\infty} dx' G(x,t+\tau|x',t) f(x',t)$. 

	Now, in many natural systems, we often observe diffusion processes, which do not follow the law in Eq.~(3) but rather the spatial spread $\lambda \propto t^{\alpha}$ with $\alpha$ different from $1/2$. Such a phenomenon is referred to as anomalous diffusion, which is of extreme general interest in contemporary statistical mechanics \cite{montroll84,bouchaud90,ben00}. Examples exhibiting anomalous diffusion are motion of tracer particles in turbulent flows \cite{richardson26}, charge transport in anomalous solids \cite{sher75}, dissolved micelles \cite{ott90}, cell migration \cite{thurner03}, chaotic dynamics \cite{shlesinger93}, porous glasses \cite{stampf95}, and subrecoil laser cooling \cite{bardou02}. To incorporate this exotic phenomenon with the framework of statistical mechanics, it seems necessary to generalize traditional kinetic theory. In fact, there are two different approaches to this problem known in the literature as fractional  \cite{zaslavsky94,appl,metzler00,zaslavsky02,grigolini99,barkai00,saxena04} and nonlinear \cite{peletier81,plastino95,tsallis96,kaniadakis02,gorban03} ones. To our knowledge, a connection between them has not been established yet on physical grounds.

	In this paper, we aim to derive and thereby unify the fractional and nonlinear theories of anomalous diffusion by appropriately generalizing Einstein's approach. 
A key point here is that the basic equation, Eq.~(1), is an integral equation. This fact allows to relax the analyticity conditions on the relevant quantities. Another point is to include nonlinearity in Eq.~(1) in conformity with the normalization condition, which may account for a nontrivial structure of the medium.  

	Let us recall that Einstein assumed the existence of the second moment of the distribution, $\phi(\Delta)$, to derive the ordinary diffusion equation. This, however, does not hold for a large class of power-law distributions, like the L\'evy stable distributions, in particular. In this case, the expansion of $f(x+\Delta, t)$ in terms of $\Delta$ does not make sense. Instead, here we note that the right-hand side of Eq.~(1) has the form of convolution of $f$ and $\phi$, which factorizes in the Fourier space:
 \begin{equation}
 \tilde f(k,t+\tau)=\tilde f(k,t) \tilde \phi(k) \quad ,
 \end{equation}  
where $\tilde f(k, t)$ and $\tilde \phi(k)$ are the characteristic functions of $f(x, t)$ and $\phi(\Delta)$, respectively, provided ${\cal F}(g)(k)=\tilde g(k) \equiv \int_{-\infty}^{\infty} dy \, g(y) e^{iky}$. Now, suppose as an example of the distribution with the divergent second moment the L\'evy distribution, whose characteristic function is given by the stretched exponential form
 \begin{equation}
 \tilde  \phi(k) = e^{-a|k|^{\gamma}} = 1- a|k|^{\gamma} + \cdots \quad ,
 \end{equation}  
where $a$ is a positive constant and the L\'evy index, $\gamma$, is in the range $(0, 2)$. $\phi(\Delta)$ decays as a power law: $\phi(\Delta) \sim |\Delta|^{-1-\gamma}$. Substituting Eq.~(5) into Eq.~(4), performing the inverse Fourier transformation, and then expanding the left-hand side in terms of $\tau$, we obtain for the leading terms the following fractional diffusion equation:
\begin{equation}
 \frac{\partial f}{\partial x}=D^* \frac{\partial^{\gamma} f}{\partial |x|^{\gamma}} \quad ,
 \end{equation}  
where the generalized diffusion constant, $D^*$, is calculated to be
\begin{equation}
D^*=  \frac{}{} \frac{a}{\tau}                                              \quad .
\end{equation}																																													 
In deriving Eq.~(6), we have used the the mathematical fact \cite{appl} that 
the fractional derivative is defined through the Fourier transformation as
 \begin{equation}
  {\cal F} \left( \frac{\partial^{\gamma} g(y)}{\partial |y|^{\gamma}} \right)(k) = -|k|^{\gamma} \tilde g(k)
 \quad .
 \end{equation} 
The solution to this equation with the singular initial condition, $f(x, 0)=\delta(x)$, is given by
 \begin{equation}
 f(x,t)=\frac{1}{2\pi} \int_{-\infty}^{\infty}  dk  e^{-ikx -D^* t|k|^{\gamma}} \quad ,                           \end{equation}																																													 
which itself is the L\'evy distribution. From the scaling law, $f(x,t)=t^{-1/\gamma} f(x/t^{1/\gamma})$, 
satisfied by the solution in Eq.~(9), the spatial spread, $\lambda$, (given by 
the half-width, for example) is seen to be, 
$\lambda \propto t^{1/\gamma}$, showing anomalous diffusion.                                                                           
It has to be emphasized that although fractional calculus has already 
been used for the discussions about L\'evy flight, the above derivation 
is the first to demonstrate its physical origin in a natural manner. 

Next, we proceed to show that Einstein's approach also provides a 
framework for diffusion phenomena in media with nontrivial structures 
with a unique viewpoint. In this case, we note that the structures can 
be taken care of by introducing nonlinearity in Eq.~(1). Thus, we consider 
the following generalization:
 \begin{equation}
 f(x,t+\tau)= \int_{-\infty}^{\infty}  d\Delta \, F(f(x+\Delta,t)) \phi(\Delta) \quad , 
 \end{equation}  
where $F(f(x, t))$ stands for the escort distribution \cite{beck93,abe03}
 \begin{equation}
 F(f(x,t))= \frac{\Phi(f(x,t))}{\int_{-\infty}^{\infty} dx' \Phi(f(x',t))}
 \end{equation}  								
with $\Phi(y)$ a positive differentiable function. 

Let us discuss a special case when the medium has the following property. 
Suppose the total medium is divided into two uncorrelated subsystems, $A$ 
and $B$, and their physical properties are identical to that of the total 
medium. In this case, the function, $\Phi$, should satisfy the following 
condition:
 \begin{equation}
 \Phi(ab)=\Phi(a)\Phi(b) \quad ,	
 \end{equation}  
which implies that $\Phi$ is a pure power-law function and correspondingly 
the escort distribution has the form	
 \begin{equation}
 F(f(x,t))= \frac{f^{\nu}(x,t)}{\int_{-\infty}^{\infty} dx' f^{\nu}(x',t)} \quad .
 \end{equation} 										
Now, assuming the existence of the second moment of $\phi(\Delta)$, and 
then following Einstein's procedure, we arrive at a family of equations
 \begin{equation}
 \frac{\partial f}{\partial t} = D^{\dagger} (t)  \frac{\partial^2 f^{\nu}}{\partial x^2} 
 + \frac{1}{\tau} \left( \frac{f^{\nu}}{c(t)}  -f \right) \quad ,
 \end{equation}  
 \begin{equation}
  D^{\dagger}(t) = \frac{D}{c(t)} \quad ,
 \end{equation}  														
where $c(t)=\int_{-\infty}^{\infty} dx f^{\nu}(x,t)$ and $D$ is given in Eq.~(2).
Except for the second term on the right-hand side as well as the time dependence 
in the generalized diffusion coefficient, $D^{\dagger}(t)$, Eq.~(14) is the 
porous medium equation \cite{peletier81}, which exhibits anomalous 
diffusion, $\lambda \propto t^{\frac{1}{1+\nu} }$ \cite{tsallis96}. 
The time dependence in the generalized 
diffusion coefficient may describe variations of the structure of the medium, 
seen from the diffusing particle.
It should be calculated by using the solution of Eq.~(14) in a self-consistent
manner. 
The second term on the right hand side can be large due to the factor, $1/\tau$.
However, the spatial integral of this term identically vanishes because of the 
normalization conditions. Therefore, it is regarded as a fluctuating contribution. 
Unfortunately, this novel equation does not seem to be analytically tractable, 
and heavy numerical analysis is yet to be carried out to examine the detailed diffusion 
properties.

In conclusion, we have revisited Einstein's  1905 theory of Brownian motion
in a modern context of anomalous diffusion. We have discussed both fractional 
and nonlinear generalizations of ordinary kinetic theory. We have shown how 
naturally fractional calculus enters into the generalized diffusion equation 
if the assumptions of analyticity and the existence of the second moment are relaxed. 
To our knowledge, 
this is the first demonstration of the physical origin of the fractional derivative, 
in marked contrast to the usual phenomenological introduction of it.
We have also generalized Einstein's approach to nonlinear kinetic theory. 
We have seen that the porous-medium-type equation can be derived if the escort 
distribution is appropriately employed.	

It is a pleasure to see how Einstein's century old idea can still provide a new 
insight into a modern aspect of Brownian motion. \\										
						
S.A. would like to thank Complex Systems Research Group of Medizinische Universit\"at 
Wien for hospitality extended to him. He was also supported in part by the Grant-in-Aid 
for Scientific Research of Japan Society for the Promotion of Science.
S.T. acknowledges support from the FWF Grant P17621.				
																																			 
\newpage

\end{document}